\documentclass[letterpaper,english,aps, pre, reprint, superscriptaddress]{revtex4-1}
\usepackage[T1]{fontenc}
\usepackage[latin9]{inputenc}
\usepackage{amsmath}
\usepackage{amssymb}
\usepackage{bbold}
\usepackage{graphicx}

\makeatletter

\pdfpageheight\paperheight
\pdfpagewidth\paperwidth

\PassOptionsToPackage{position=top}{subfig}
\usepackage{hyperref}
\hypersetup{
	colorlinks = true,
	allcolors = blue
}

\usepackage{times}

\makeatother

\usepackage{babel}

\begin{document}

\title{Predicted critical state based on invariance of the Lyapunov exponent in dual spaces}
\author{Tong Liu}
%\thanks{t6tong@njupt.edu.cn}
\affiliation{Department of Applied Physics, School of Science, Nanjing University of Posts and Telecommunications, Nanjing 210003, China}
\author{Xu Xia}
\thanks{xiaxu14@mails.ucas.ac.cn}
\affiliation{Academy of Mathematics and System Sciences, Chinese Academy of Sciences, Beijing 100190, China}

\date{\today}

\begin{abstract}
The critical state in disordered systems, a fascinating and subtle eigenstate, has attracted a lot of research interest. However, the nature of the critical state is difficult to describe quantitatively, and in general it cannot predict a system that host the critical state. In this work, we propose an explicit criterion that Lyapunov exponent of the critical state should be 0 simultaneously in dual spaces, namely Lyapunov exponent remains invariant under Fourier transform. With this criterion, we exactly predict a system hosting a large number of critical states. Then, we perform numerical verification of the theoretical prediction, and display the self-similarity of the critical state. Finally, we conjecture that there exists some kind of connection between the invariance of the Lyapunov exponent and conformal invariance, which can promote the research of critical phenomena.
\end{abstract}

%\pacs{71.23.An, 71.23.Ft, 05.70.Jk}
\maketitle

\section{Introduction}
Since the phenomenon of Anderson localization~\cite{Anderson} was proposed, the quantum disordered system has attracted extensive research enthusiasm. When impurities are randomly added to an ideal conductor, Bloch state in the conductor will change into localized state with the increase of the average concentration of impurities. Heretofore many landmark achievements have been obtained~\cite{scaling1,scaling2}. For a three-dimensional tight-binding model with short-range hopping and random on-site potential, there exists a critical energy separating extended state and localized state, which is dubbed mobility edge~\cite{Mott}. And many rigorous mathematical methods were developed to quantitatively analyze Anderson localization phenomenon~\cite{A50}. An example is to use the transfer matrix method to solve the Lyapunov exponent of one-dimensional Anderson model~\cite{MA}, as one-dimensional problem is more likely to be solved analytically.

Transfer matrix method focuses on the recurrence relation of the amplitude of wave functions, and is to calculate the average exponential divergence rate ~\cite{Soukoulis}.
Mathematically, let Lyapunov exponent $\gamma$ be rewritten as the form of $\exp(-\gamma)$, the latter indicates the divergence rate between adjacent lattice points. In position space, for an extended state, the amplitude between neighboring lattice points should be equal under the thermodynamic limit, so $\gamma$ should be 0; whereas for a localized state, the amplitude between neighboring lattice points should be exponentially decaying, so $\gamma$ should be greater than 0. Consequently, we can utilize the Lyapunov exponent to explicitly distinguish the state, namely, the extended state corresponds to $\gamma=0$ and localized state corresponds to $\gamma>0$.

However, in the quantum disordered system, there is a class of rare and important states, namely the critical state, the rarity means that it generally only appears at the phase transition point~\cite{Sokoloff}, such as the eigenstate at the mobility edge. The critical state is neither an extended state nor a localized state, its prominent feature is possessing the self-similar structure~\cite{A50}. In the language of multifractal theory~\cite{fractal}, the minimum scaling index of the extended state should tend to 1, that of the localized state should tend to 0, and that of the critical state should be greater than 0 but less than 1. However, Multifractal theory requires that the system should be numerically diagonalized firstly, then using numerically obtained eigenstates to calculate the minimum scaling index. This method cannot predict the nature of the eigenstate before numerically diagonalizing the Hamitonian of a system.

In contrast, calculating the Lyapunov exponent does not need detailed information about eigenstates, and it can directly be solved by giving the Hamiltonian~\cite{Avila1}. As long as Lyapunov exponent of the system being obtained, we can predict the properties of the eigenstates without numerical diagonalization. For the famous Aubry-Andr\'{e} model~\cite{AA}, Lyapunov exponent is $\gamma=\max\{0,\ln(V/2)\}$ in position space, and $\gamma_m=\max\{0,\ln(2/V)\}$ in momentum space, where $V$ is the strength of quasiperiodic potential. We can accurately predict when
$0<V<2$, $\gamma=0$, the eigenstate in position space is extended; while $V>2$, $\gamma>0$, the eigenstate is localized. However, a confusion emerges, when $V=2$ (the phase transition point), $\gamma$ is also equal to 0, at this point the eigenstates are all critical states. Consequently, the Lyapunov exponent cannot distinguish the extended state from the critical state, the $\gamma=0$ method is invalid.

Therefore, numerous research efforts are devoted to accurately describing the properties of critical states. Reference~\cite{Goncalves} develops a renormalization-group theory to describe the localization properties of quasiperiodic systems. This theory can be used to get exact or approximate analytical expressions for the phase boundaries of extended, localized and critical phases for specific models. Reference~\cite{GM} proposes that the coupling between localized and extended states in their overlapped spectrum can provide a general recipe to construct critical states for two-chain models. Reference~\cite{LXJ} points out that there are two approaches to involve the existence of critical states, one is involving unbounded potential~\cite{LT} and the other is involving zeros of hopping terms in Hamiltonian. However, an explicit criterion for characterizing critical states is still lacking.

To fill this theoretical gap, we give a quantitative criterion to predict the critical state in general sense. Let's re-examine Lyapunov exponent in both
position space and momentum space, for extended state, we have $\gamma=0$ and $\gamma_m>0$; for localized state, we have $\gamma>0$ and $\gamma_m=0$. In order to distinguish from the above two states while maintaining the approaching 0 property of Lyapunov exponent, a reasonable conjecture is that the critical state corresponds to the invariance of Lyapunov exponent in dual spaces, namely its distinguishing feature is satisfying the condition
\begin{equation}\label{eq0}
\gamma=\gamma_m=0.
\end{equation}
To verify the validity of this conjecture, we introduce an exactly solvable model hosting critical states in a wide range of parameters, and exactly predict the interval of the existence of critical states through $\gamma=\gamma_m=0$. Hence we demonstrate that this conjecture is applicable to the various critical states.

\section{Model and prediction}
The difference equation of the model that we consider is written as,
\begin{equation}\label{eq1}
\psi_{n+1}+\psi_{n-1}+V i \tan(2\pi\alpha n+\theta) \psi_n = E \psi_n,
\end{equation}
where $V$ is the complex potential strength, $E$ is the eigenvalue of systems, and $\psi_n$ is the amplitude of wave function at the $n$th lattice. We choose to unitize the nearest-neighbor hopping amplitude, a typical choice for irrational parameter is $\alpha=(\sqrt{5}-1)/2$, and $\theta$ is the phase factor.

With the Hamiltonian of the system [Eq.~(\ref{eq1})], we can determine Lyapunov exponent $\gamma$ in position space, which is calculated by taking the product of the transfer matrix $T(\theta)$, namely $\gamma = \lim_{n\rightarrow\infty}\ln||T_n(\theta)||/n.$
Utilizing Avila's global theory~\cite{Avila}, we can get the explicit expression of $\gamma$, see detailed calculation in Supplemental Material~\cite{Supplemental},
\begin{equation}\label{eq2}
\begin{aligned}
\gamma(E) = \max\{&\operatorname{arcosh} \frac{\left| E+V+2\right|+\left| E+V-2\right|}{4},\\
&\operatorname{arcosh} \frac{\left|E-V+2\right|+\left|E-V-2\right|}{4}\} .
\end{aligned}
\end{equation}
From Eq.~(\ref{eq2}), we can extract the allowed energies of the system. Firstly, we make $\gamma(E)=0$, and obtain that $E$ is within the region $ [V-2, 2-V]$. Secondly, we make $\gamma(E)>0$, and obtain that $E$ is within the region $\{i y~|~y\in \mathbb{R}^{*}\}$ ($V\leq 2$) or $\{i y~|~y\in \mathbb{R}\}$ ($V > 2$), which means that when the eigenstate is a localized state in the position space, the eigenvalue is a pure imaginary number.
However, when $\gamma=0$, with regard to the eigenvalue within $[V-2, 2-V]$, it cannot be concluded that the corresponding eigenstate is an extended state or a critical state.

The next step is to calculate Lyapunov exponent $\gamma_m$ in momentum space. Firstly, we introduce the Fourier transformation
$$
f_k=\frac{1}{\sqrt{L}} \sum_{n=1}^L e^{i 2\pi \alpha k n} \psi_n,
$$
thus the dual equation of Eq.~(\ref{eq1}) in momentum space is written as
\begin{equation}\label{eq3}
{f_{k+1}} =\frac{-2\cos [2 \pi (k-1) \alpha]+V +E}{2\cos [2 \pi (k+1)\alpha]+V -E } f_{k-1}.
\end{equation}

From Eq.~(\ref{eq3}), an initial wave function solution can be written as
$$
f_{k} \propto\left\{\begin{array}{cl}
0, & k=0,\pm 2,\pm 4 \cdots, \\
0, & k=2 j+1<k_0, \\
1, & k=2 j +1=k_0, \\
\frac{g^{(1)}_{k-2}}{g^{(2)}_{k}} f_{k-2}, & k=2 j+1>k_0,
\end{array}\right.
$$
%Such a solution is an eigenstate provided that $\left|f_{k}\right|$ does not diverge as $k \rightarrow$ $\infty$.
%In particular, if $f_{k} \rightarrow 0$ as $k \rightarrow \infty$ and $\sum_n\left|f_{k}\right|^2<\infty$, the eigenstate is localized.
then Lyapunov exponent $\gamma_m$ can be obtained by Sarnark's method~\cite{Supplemental,Sarnak},
\begin{equation}
\gamma_m(E)=\lim _{k \rightarrow \infty} \frac{1}{k-k_0} \ln \left|\frac{f_{k}}{f_{k_0}}\right|=\frac{1}{2 \pi} \int_0^{2 \pi}\ln g^{(1)}-\ln g^{(2)} d \theta,
\end{equation}
where $g^{(1)}=|-2\cos (2 \pi  \theta)+V +E |,g^{(2)}=|2\cos (2 \pi  \theta)+V -E |$.

Recalling the calculation process of Lyapunov exponent $\gamma$ in position space, when $\gamma>0$, namely eigenvalues of the system are pure imaginary numbers, we have the identity
\begin{equation}\label{eq51}
\begin{aligned}
&\frac{1}{2 \pi} \int_0^{2 \pi} \ln |-2\cos (2 \pi  \theta)+V  +i y | d \theta\\
&=\frac{1}{2 \pi} \int_0^{2 \pi} \ln |2\cos (2 \pi  \theta)+V -i y | d \theta,
\end{aligned}
\end{equation}
hence Lyapunov exponent in momentum space $\gamma_m=0$. $\gamma>0$ and $\gamma_m=0$ indicate the associated state is the extended state in momentum space, which is exactly corresponding to the localized state in position space.

When $\gamma=0$, namely eigenvalues are real numbers and within the interval $[V-2, 2-V]$, we have
\begin{equation}\label{eq5}
\begin{aligned}
\frac{1}{2 \pi} \int_0^{2 \pi} \ln|g^{(1)}| d\theta=
\left\{\begin{array}{cl}0, & g^{(1)}=0 ,  \\ \ln \left|\frac{|E+V| +\sqrt{(E+V)^2-4}}{2}\right|, & g^{(1)}\neq 0, \end{array}\right.
\end{aligned}
\end{equation}
and
\begin{equation}\label{eq6}
\begin{aligned}
\frac{1}{2 \pi} \int_0^{2 \pi} \ln|g^{(2)}| d\theta=
\left\{\begin{array}{cl}0, & g^{(2)}=0,  \\ \ln \left|\frac{|E-V| +\sqrt{(E-V)^2-4}}{2}\right|, & g^{(2)}\neq 0. \end{array}\right.
\end{aligned}
 \end{equation}
According to Eq.~(\ref{eq5}) and Eq.~(\ref{eq6}), we obtain the identity $$\frac{1}{2 \pi} \int_0^{2 \pi} \ln|g^{(1)}| d\theta=\frac{1}{2 \pi} \int_0^{2 \pi} \ln|g^{(2)}| d\theta,$$see detailed calculation in Supplemental Material~\cite{Supplemental}.
Hence, when $V-2\leq E \leq 2-V$, $\gamma_m$ is also equal to 0, which is very similar to the result that the eigenvalues are pure imaginary numbers. The difference is that the latter have $\gamma>0$ in position space and $\gamma_m=0$ in momentum space; whereas the eigenstates of the former have $\gamma=\gamma_m=0$.

Based on the above achievements and the conjecture Eq.~(\ref{eq0}), we can predict that different from the common models, where the critical state only exists at the phase transition point, for the model of Eq.~(\ref{eq1}), critical states exist in a wide range of the parameters $0<V\leq2$, due to $V-2\leq E \leq 2-V$ indicating $0<V\leq2$. Consequently, we realized the prediction of a system hosting a large number of critical states through giving its Hamiltonian.

\section{Numerical verification and Self-similarity}
\begin{figure}
  \centering
  \includegraphics[width=0.48\textwidth]{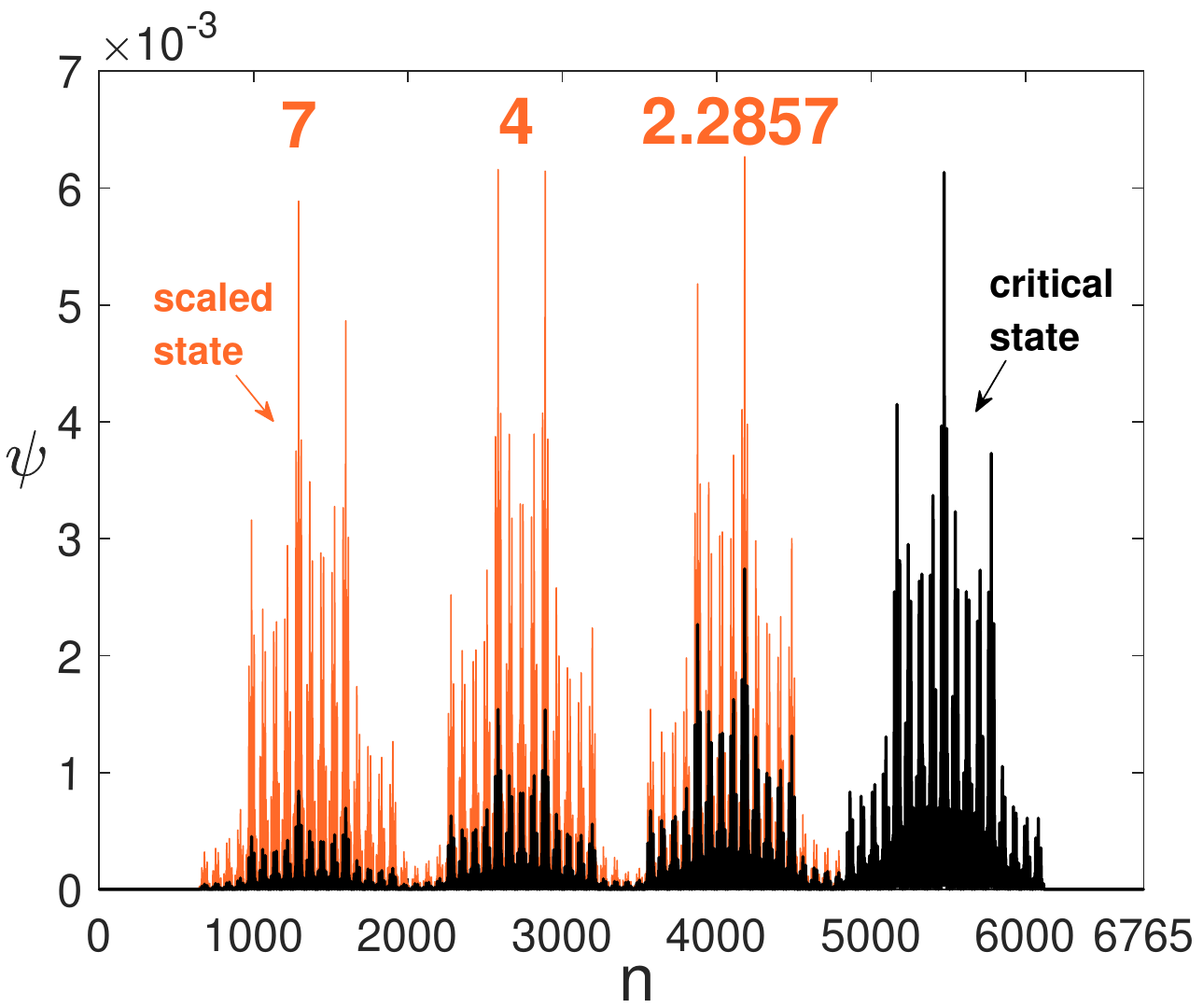}\\
  \caption{(Color online) The black curve represents wave function of $E=-1$ obtained from Eq.~(\ref{eq1}) with the parameter $V=1$. The red curves represent three wave function peaks after magnifying. Starting with the smallest peak, the scaled multiples are 7, 4 and 2.2857 in turn. It clearly shows that the scaled three smaller peaks are very similar to the largest peak. The total number of sites is set to be $L=6765$.}
\label{fig1}
\end{figure}
\begin{figure}%[H]
  \centering
  \includegraphics[width=0.48\textwidth]{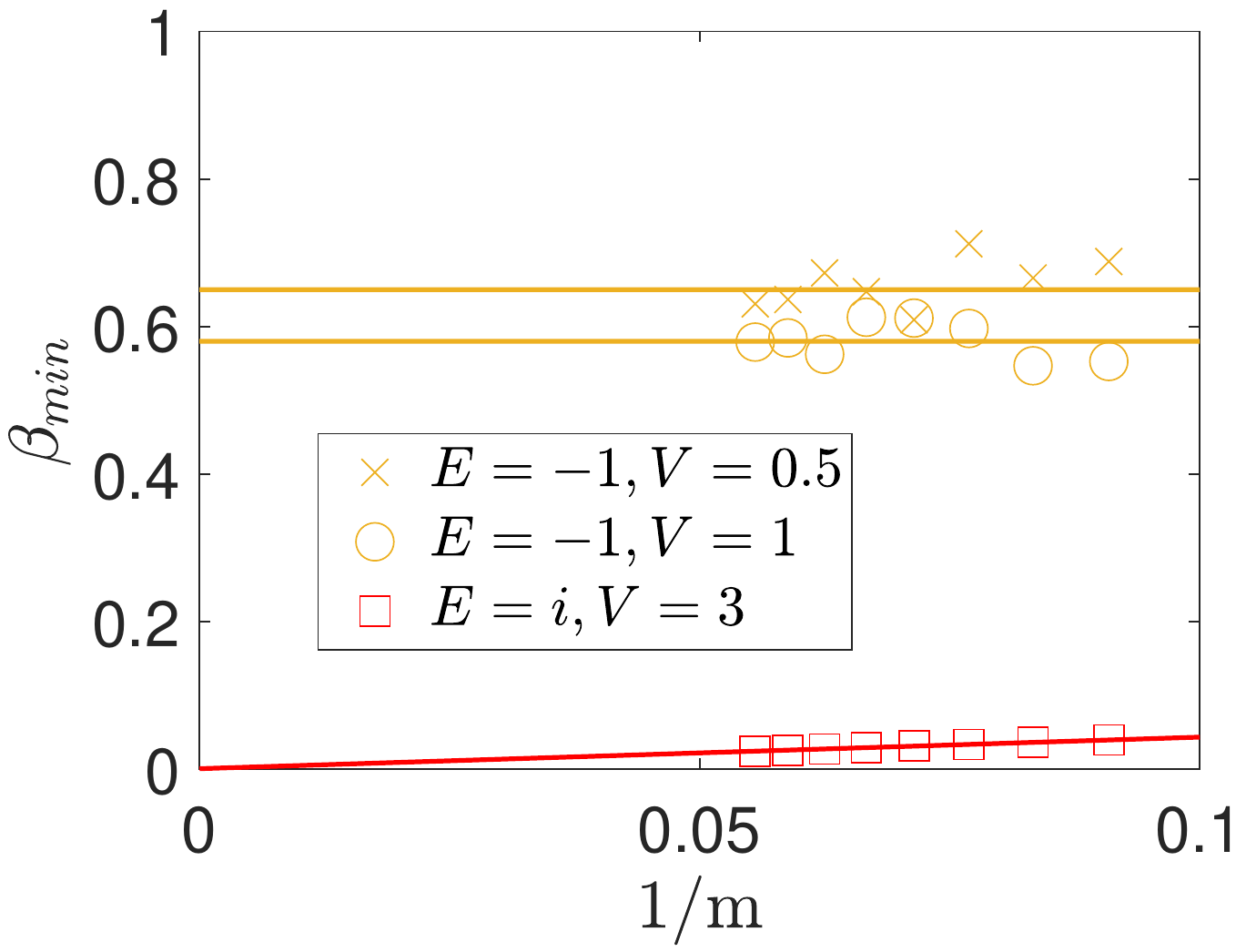}\\
  \caption{(Color online) $\beta_{min}$ as a function of the inverse Fibonacci index $1/m$ for various eigenvalues. In the large size limit, the brown "x" markers tend to 0.67 and correspond to the critical state with $E=-1$ ($V=0.5$), the brown "o" markers tend to 0.58 and correspond to the extended state with $E=-1$ ($V=1$), the red square markers tend to 0 and correspond to the localized state with $E=i$ ($V=3$).
  }
  \label{f2}
\end{figure}

To support the analytical results given in the previous section, we perform numerical calculations and analyze the critical nature of the eigenstate. We numerically diagonalize the Hamiltonian~(\ref{eq1}) in a large size, and get the eigenvalue and the associated eigenstates. To verify the critical state, we should illustrate the distinctive feature of the corresponding state, namely self-similarity.

Mathematically, self-similarity is a typical property of the fractal~\cite{fractal1,fractal2}, a self-similar construction is exactly or approximately similar to a part of itself. Many physical objects in nature, such as capillary distribution and leaf veins, are statistically self-similar: parts of them show the same statistical properties as the whole. An equivalent description of self-similarity is scale invariant~\cite{fractal3,fractal4}, where there is a smaller part that similar to the proximate larger part at the certain amplitude. 
Thus, we can perform the contraction-expansion variations on the part of physical quantities of the system. If the variation amplitude meets a certain value, it means that the physical quantity is scale invariant, that is, the system has a self-similarity structure. 
%The associated scaling transformation refers to the contraction-expansion variations of a certain region of the space or the time. When the physical quantity of the system before and after the scaling transformation meets a certain magnification, the system has scale invariance. 
%For instance, a side of the Koch snowflake is scale-invariant after scaling transformation, which can be continuously expanded by a certain multiple, whereas its shape looks unchanged.

Therefore, we directly numerically diagonalize Eq.~(\ref{eq1}), plot the obtained eigenstate, and examine detailed structure of wave function to determine whether there exists self-similarity. As shown in Fig.~\ref{fig1}, the black curve is the original eigenstate of the system, it has four wave-function peaks of different amplitudes. This state is not a localized state, but it does not look like an extended state, because the peaks of the extended state should be equal. When we magnify three smaller peaks by 7, 4 and 2.2857 times successively, namely the red curves, they become very similar to the largest black peak. More importantly, we found that $7/4\approx4/2.2857\approx1.75$, which indicates that three smaller peaks can be expanded to the largest one by a certain multiple $1.75$. This clearly shows that wave function of the system has scale invariance and self similarity, hence this state is definitely a critical state. We also calculate the eigenstates corresponding to different energy levels and different sizes, and all numerical results are critical states as expected, see detailed calculation in Supplemental Material~\cite{Supplemental}.
For the critical state of known models~\cite{known1,known2}, the validity of the criterion can also be conveniently verified.

In addition to visually displaying the wave function, we also calculate the minimum scaling index of the critical state according the multifractal theory~\cite{fractal}. 
For giving the wave function $\psi_n$, a scaling index $\beta_{n}$ can be extracted from the $n$th
on-site probability $P_{n} = \vert\psi_n \vert^2 \sim (1/F_{m})^{\beta_{n}}$, where $F_{m}$ is the $m$th Fibonacci number.
The multifractal theorem states, when the wave functions are extended, the maximum of $P_{n}$ scales as $\max(P_{n})
\sim (1/F_{m})^1$, i.e., $\beta_{min}=\min(\beta_{n})=1$. On the other hand, when the wave functions are localized, $P_{n}$ concentrates at the individual site and tends to zero at the other sites, yielding $\max(P_{n}) \sim (1/F_{m})^0$ and $\beta_{min}=\min(\beta_{n})=0$. With regard to the critical state, the corresponding $\beta_{min}$ is located within the interval $\left(0,~1\right)$, and can be utilized to distinguish extended and critical states.
In order to reduce finite-size effects, we examines the trend of $\beta_{min}$ under the limit of large size.

As shown in Fig.~\ref{f2}, $\beta_{min}$ is plotted as a function of the inverse Fibonacci index $1/m$, when $m\rightarrow0$, the system size  $L\rightarrow\infty$. It clearly shows that $\beta_{min}$ is between $0$ and $1$ in the large $L$ limit for the eigenvalues $E=-1$ ($V=0.5$) and $E=-1$ ($V=1$), hence the corresponding state is critical.
While for the eigenenergy $E=i$ ($V=3$), $\beta_{min}$ asymptotically tends to 0 in the large $L$ limit, indicating that the corresponding state is localized. We have also checked other combinations of parameters and get the same results as expected. Thus, above numerical results are in excellent agreement with the analytical results.

\section{Conclusion} 
In this work, we propose a conjecture to predict and identify the existence of the critical state in a system, that is, Lyapunov exponent of the eigenstate should be 0 in both position space and momentum space. To illustrate this criterion, we introduce an exactly solvable model, predict and verify the existence of a large number of critical states in the wide range of the potential strength. This demonstrates that $\gamma=\gamma_m=0$ is not limited to the critical state at the phase transition point, but is applicable to various types of critical states. Our findings provide an explicit quantitative description of the characteristics of the critical state, and have a positive significance for exploring the critical state of unknown systems.
 
Furthermore, it is noteworthy that the prominent feature of critical states is the scale invariance of wave
functions, which indicates that the invariance of Lyapunov exponent under Fourier transform has a closed relation with conformal invariance. This is likely to be a new application scenario of conformal invariance theory, which can describe properties of quantum disordered systems near the critical point or within the critical interval.

%%%%%%%%%%%%%%%%
\begin{acknowledgments}
T. L. thanks Ming Gong for beneficial comments.
This work was supported by the Natural Science Foundation of Jiangsu Province (Grant No. BK20200737), NUPTSF
(Grants No. NY220090 and No. NY220208), the Innovation Research Project of Jiangsu Province (Grant No. JSSCBS20210521), and China Postdoctoral Science Foundation (Grant No. 2022M721693).
\end{acknowledgments}

%%%%%%%%%%%%%%%%%%%%%%%

%%%%%%%%%%%%%%%%%%%%%%%%%%%%%%%%%%%%%% %%   Supplementary Information %%%%%%%%%%%%%%%%%%%%%%%%%%%%%%%%%%%%%%
\renewcommand{\thesection}{S-\arabic{section}}
\setcounter{section}{0}  %  this will re-count section from 1
\renewcommand{\theequation}{S\arabic{equation}}
\setcounter{equation}{0}  %  this will re-count eq from 1
\renewcommand{\thefigure}{S\arabic{figure}}
\setcounter{figure}{0}  %  this will re-count eq from 1
\renewcommand{\thetable}{S\Roman{table}}
\setcounter{table}{0}  %  this will re-count eq from 1
\onecolumngrid \flushbottom %\onecolumn

\newpage
\begin{center}\large \textbf{Supplementary Material} \end{center}
This Supplemental Material provides additional information for the main text. In Sec.~\ref{position}, we provide the details of calculating Lyapunov exponent in position space. In Sec.~\ref{momentum}, we provide the details of calculating Lyapunov exponent in momentum space. Finally, we give more numerical verification of critical state  in Sec.~\ref{numerical}.

\section{Derivation of Lyapunov exponent $\gamma$ in position space}\label{position}
The Lyapunov exponent $\gamma$ can be calculated by taking the product of the transfer matrix $T(\theta)$, namely multiplying the transfer matrix $n$ times consecutively, which is written as $$T_n(\theta)=\prod_{l=0}^{n-1}T(2\pi\alpha l+\theta)=\prod_{l=0}^{ n-1}\left(\begin{array}{cc} E-V i\tan(2\pi\alpha l+\theta) & -1 \\ 1 & 0\end{array}\right) ,$$then Lyapunov exponent is $\ln||T_n(\theta)||/n$ as $n$ tends to the infinite in the thermodynamic limit.

The method we use here to calculate Lyapunov exponent is the complexified phase approach, specifically by continuing the imaginary part of the phase $\epsilon$, we focus on the new Lyapunov exponent, that is

$$T_n(\theta+ i\epsilon)=\prod_{l=0}^{n-1}T(2\pi\alpha l+\theta+ i\epsilon),$$ correspondingly, we get $\gamma(\epsilon)$ is $\lim_{n\rightarrow\infty}\ln||T_n(\theta+i\epsilon)||/n.$

Relying on Avila's global theory ~\cite{Avila2015}, if we can obtain Lyapunov exponent $\gamma(\epsilon)$ when $\epsilon$ is sufficiently large, then we can trace back to the specific Lyapunov exponent $\gamma(0)$ when $\epsilon=0$, namely the original Lyapunov exponent $\gamma$ in position space.

Firstly, rewriting the transfer matrix
\begin{equation}\label{seq1}
\begin{aligned}
T(\theta)&= \left(\begin{array}{cc} E-V i\tan(2\pi\alpha l+\theta) & -1 \\ 1 & 0\end{array}\right) \\
&=\sec(2\pi\alpha l+\theta)B(\theta)
\end{aligned}
 \end{equation}
where

\begin{equation}\label{seq2}
\begin{aligned}
 B(\theta)=\left(\begin{array}{cc} E\cos(2\pi\alpha l+\theta)-V i \sin(2\pi\alpha l+\theta) & -\cos(2\pi\alpha l+\theta) \\ \cos(2\pi\alpha l+\theta) & 0\end{array}\right)
\end{aligned}
 \end{equation}
then, $\gamma(\epsilon)$ can be expressed as
\begin{equation}\label{seq3}
\begin{aligned}
\gamma_{\epsilon}(E)=&\lim_{n\rightarrow\infty}\ln||T_n(\theta+i\epsilon)||/n\\
=&\lim_{n\rightarrow \infty}\frac{1}{n} \int\ln|B_n(\theta +i\epsilon )| d\theta + \int\ln |\sec(\theta +i\epsilon)| d\theta.\\
=&\gamma_{\epsilon}^B(E)+\ln(2)-2\pi\epsilon,
\end{aligned}
\end{equation}
where
$$
\begin{aligned}
&\gamma_{\epsilon}^B(E)=\lim _{n \rightarrow \infty} \frac{1}{n} \int \ln \left|B_n(\theta+i \varepsilon)\right| d \theta, \\
&B_n(\theta+i c)=\prod_{l=0}^{n-1} B(2 \pi \alpha l+\theta+i \epsilon).
\end{aligned}
$$

When $\epsilon$ tends to  $+\infty$, a direct calculating result of $B(\theta+i\epsilon)$ is
\begin{equation}\label{seq4}
B(\theta+i\epsilon)=-\frac{1}{2} e^{2\pi\epsilon}e^{i2\pi(\theta+\alpha)}\left(
\begin{array}{cc}
E-V & -1 \\\\
1 & 0 \\\\
\end{array}
\right) + o(1).
\end{equation}
Thus we get ${\gamma^{+B}}_{\epsilon}(E)=2\pi\epsilon+\ln|
\frac{\sqrt{(E-V)^2\pm 4}}{2}| -\ln (2)+o(1)$.
As a function of $\epsilon$, $\gamma_{\epsilon}^B(E)$ is a convex, piecewise linear function whose slope is an integer multiplied by $2\pi$, hence it is concluded that when $\epsilon$ tends to infinity, we obtain${\gamma^{+B}}_{\epsilon}(E)=2\pi\epsilon+\ln|\frac{\sqrt{(E-V)^2\pm 4}}{2}| -\ln (2)$.
And according to the equation(\ref{seq3}), it leads that when $\epsilon$ is the very large positive number,
${\gamma^{+}}_{\epsilon}(E)
={\gamma^{+B}}_{\epsilon}(E)+\ln(2)-2\pi\epsilon=\ln|\frac{\sqrt{(E-V)^2\pm 4}}{2}|.
$

When $\epsilon$ tends to  $-\infty$, a direct calculating result of $B(\theta+i\epsilon)$ is
\begin{equation}\label{seq4}
B(\theta+i\epsilon)=-\frac{1}{2} e^{2\pi\epsilon}e^{i2\pi(\theta+\alpha)}\left(
\begin{array}{cc}
E+V & -1 \\\\
1 & 0 \\\\
\end{array}
\right) + o(1).
\end{equation}
Thus we get ${\gamma^{-B}}_{\epsilon}(E)=2\pi\epsilon+\ln|
\frac{\sqrt{(E+V)^2\pm 4}}{2}| -\ln (2)+o(1)$.
As a function of $\epsilon$, $\gamma_{\epsilon}^B(E)$ is a convex, piecewise linear function whose slope is an integer multiplied by $2\pi$, hence it is concluded that when $\epsilon$ tends to infinity, we obtain ${\gamma^{-B}}_{\epsilon}(E)=2\pi\epsilon+\ln|
\frac{\sqrt{(E+V)^2\pm 4}}{2}| -\ln (2)$.
And according to equation(\ref{seq3}), it leads that when $\epsilon$ is the very large negative number,
${\gamma^{-}}_{\epsilon}(E)
={\gamma^{-B}}_{\epsilon}(E)+\ln(2)-2\pi\epsilon=\ln|\frac{\sqrt{(E+V)^2\pm 4}}{2}|.
$

Since $\gamma_{\epsilon}$ is a convex function in two semilinear $(0,+\infty)$ and $(-\infty, 0)$, it is linear in the cross section and the slope is an integer multiplied by the $2\pi$ integer, Lyapunov exponent is $$
\gamma_{\varepsilon}(E)=\left\{\begin{array}{cc}
\gamma_{\varepsilon}^{+}(E) & \varepsilon>0, \\
\gamma_{\varepsilon}^{+}(E)+2 \varepsilon & \frac{\gamma_{\varepsilon}^{-}(E)-\gamma_{\varepsilon}^{+} (E)}{2}<\varepsilon<0, \\
\gamma_{\varepsilon}^{-}(E)  & \varepsilon<\frac{\gamma_{\varepsilon}^{-}(E)-\gamma_{\varepsilon}^{+} (E)}{2},\end{array}\right.
$$ the relationship between the left and right limit conditions is $\gamma_{\varepsilon}^{+}(E)>\gamma_{\varepsilon}^{-}(E)$ for any given value of $\varepsilon$.

Similarly, if the large and small relationship between the left and right limits is $\gamma_{\varepsilon}^{+}(E)<\gamma_{\varepsilon}^{-}(E)$, Lyapunov exponent is $$
\text { then } \gamma_{\varepsilon}(E)=\left\{\begin{array}{cc}
\gamma_{\varepsilon}^{-}(E) & \varepsilon<0, \\
\gamma_{\varepsilon}^{-}(E)-2 \varepsilon & 0<\varepsilon<\frac{\gamma_{\varepsilon}^{-}(E)-\gamma_{\varepsilon}^{+} (E)}{2}, \\
\gamma_{\varepsilon}^{+}(E)  & \varepsilon>\frac{\gamma_{\varepsilon}^{-}(E)-\gamma_{\varepsilon}^{+} (E)}{2}.\end{array}\right.
$$

Summarizing the above conclusions, Lyapunov exponent in position space is $\gamma=\max\{\gamma_{\varepsilon}^{+}(E), \gamma_{\varepsilon}^{-}(E)\}$, which is
\begin{equation}\label{seq6}
\begin{array}{r}
\gamma(E)=\max \left\{\operatorname{arcosh} \frac{|E+V+2|+|E+V-2|}{4},\right. \\
\left.\operatorname{arcosh} \frac{|E-V+2|+|E-V-2|}{4}\right\} .
\end{array}
\end{equation}

\section{Derivation of Lyapunov exponent $\gamma_m$ in momentum space}\label{momentum}
In the main text, utilizing Fourier transform, the initial wave function solution has been obtained. Relying on Jensen's formula~\cite{Jensen}, our calculation supposes that $f$ is an analytic function, $a_1, a_2, \ldots, a_n$ are the zeros of $f$ in the interior of the unit disc of the complex plane, and $f(0) \neq 0$. Then, we have the following equality
\begin{equation}\label{jesen}
    \ln |f|=\sum_{k=1}^n \ln \left({\left|a_k\right|}\right)+\frac{1}{2 \pi} \int_0^{2 \pi} \ln \left|f\left( e^{i \theta}\right)\right| d \theta.
\end{equation}
Combining the initial wave function solution, the expression of Lyapunov exponent can be written as
$$
\gamma_m(E)=\lim _{k \rightarrow \infty} \frac{1}{k-k_0} \ln \left|\frac{f_k}{f_{k_0}}\right|=\int \ln g^{(1)}-\ln g^{(2)} d \theta,
$$
where $g^{(1)}=|-2 \cos (2 \pi \theta)+V+E|$, $g^{(2)}=\mid 2 \cos (2 \pi \theta)+$ $V-E \mid$.

Then, the first term of the rightmost side of $\gamma_{m}$ can be written as
\begin{equation}
\begin{aligned}
  \int_{0}^{1} \ln g^{(1)} d \theta&=\int_{0}^{1} \ln |-2 \cos (2 \pi \theta)+V+E| d \theta  \\
  &=\int_{0}^{1} \ln |-e^ {i 2 \pi \theta}-e^ {-i 2 \pi \theta}+V+E| d \theta\\
  %=\frac{1}{2 \pi}\int_{0}^{2 \pi} \ln |-e^ { \theta}-e^ {- \theta}+V+E| d \theta \\
  &=\frac{1}{2 \pi}\int_{0}^{2 \pi} \ln |e^ {i 2 \theta}+1-(V+E)e^ {i \theta}| d \theta.
\end{aligned}
\end{equation}
Applying Jensen's formula~\cite{Jensen}, the integral calculation of Eq.~S8 can be transformed to the calculation of roots of Eq.~S9 in the unit disc, let $x=e^ {i \theta}$,
\begin{equation}
x^2+1-(V+E)x =0.
\end{equation}
After some mathematical calculations, we obtain
\begin{equation}
\left\{\begin{array}{l}
\text { when } |V+E|<2, \int_{0}^{1} \ln g^{(1)} d \theta=0,\\
\text { when } |V+E|>2, \int_{0}^{1} \ln g^{(1)} d \theta=\ln\frac{|E+V|+\sqrt{(E+V)^2-4}}{2}.
\end{array}\right.
\end{equation}
Under the similar process, with regard to the second term of $\gamma_{m}$, we also obtain
\begin{equation}
\left\{\begin{array}{l}
\text { when } |V-E|<2, \int_{0}^{1} \ln g^{(2)} d \theta=0,\\
\text { when } |V-E|>2, \int_{0}^{1} \ln g^{(2)} d \theta=\ln\frac{|E-V|+\sqrt{(E-V)^2-4}}{2}.
\end{array}\right.
\end{equation}

Interesting, when both $|V+E|<2$ and $|V-E|<2$ hold, we have
\begin{equation}
\gamma_m(E)=\int \ln g^{(1)}-\ln g^{(2)} d \theta=0.
\end{equation}
When both $|V+E|<2$ and $|V-E|>2$ hold, then
\begin{equation}
\gamma_m(E)=\int \ln g^{(1)}-\ln g^{(2)} d \theta=0-\ln\frac{|E-V|+\sqrt{(E-V)^2-4}}{2}<0.
\end{equation}
When both $|V+E|>2$ and $|V-E|<2$ hold, then
\begin{equation}
\gamma_m(E)=\int \ln g^{(1)}-\ln g^{(2)} d \theta=\ln\frac{|E+V|+\sqrt{(E+V)^2-4}}{2}-0>0.
\end{equation}
When both $|V+E|>2$ and $|V-E|>2$ hold, then
\begin{equation}
\gamma_m(E)=\int \ln g^{(1)}-\ln g^{(2)} d \theta=\ln\frac{|E+V|+\sqrt{(E+V)^2-4}}{2}-\ln\frac{|E-V|+\sqrt{(E-V)^2-4}}{2}\neq 0.
\end{equation}

To sum up the above calculations, only when both $|V+E|<2$ and $|V-E|<2$ hold, namely the eigenvalues $E \in [V-2, 2-V]$, $\gamma_m$ is equal to 0.

\newpage

\section{More numerical verification}\label{numerical}
In this section, we provide more numerical validation to strengthen the credibility of our theoretical prediction. We have calculated the wave functions of different energy levels at the same size, and the same energy level at different sizes. The numerical results are as expected, and the corresponding eigenstates dispaly self-similarity. As shown in Fig. S1 and Fig. S2, it clearly illusrates that the scaled smaller peaks are very similar to the largest peak. More interesting, in Fig. S2, three left small peaks satisfy the scale invariance with a certain multiple $5.7/3.2\approx3.2/1.8\approx1.78$; whereas three right small peaks satisfy the scale invariance with a certain multiple $13.9/5.7\approx34/13.9\approx2.44$. The different scaling factors indicate that there is more than one fractal structure in the wave function, which exactly corresponds to the multifractal theory in the introduction.
\begin{figure}
  \centering
  \includegraphics[width=0.5\textwidth]{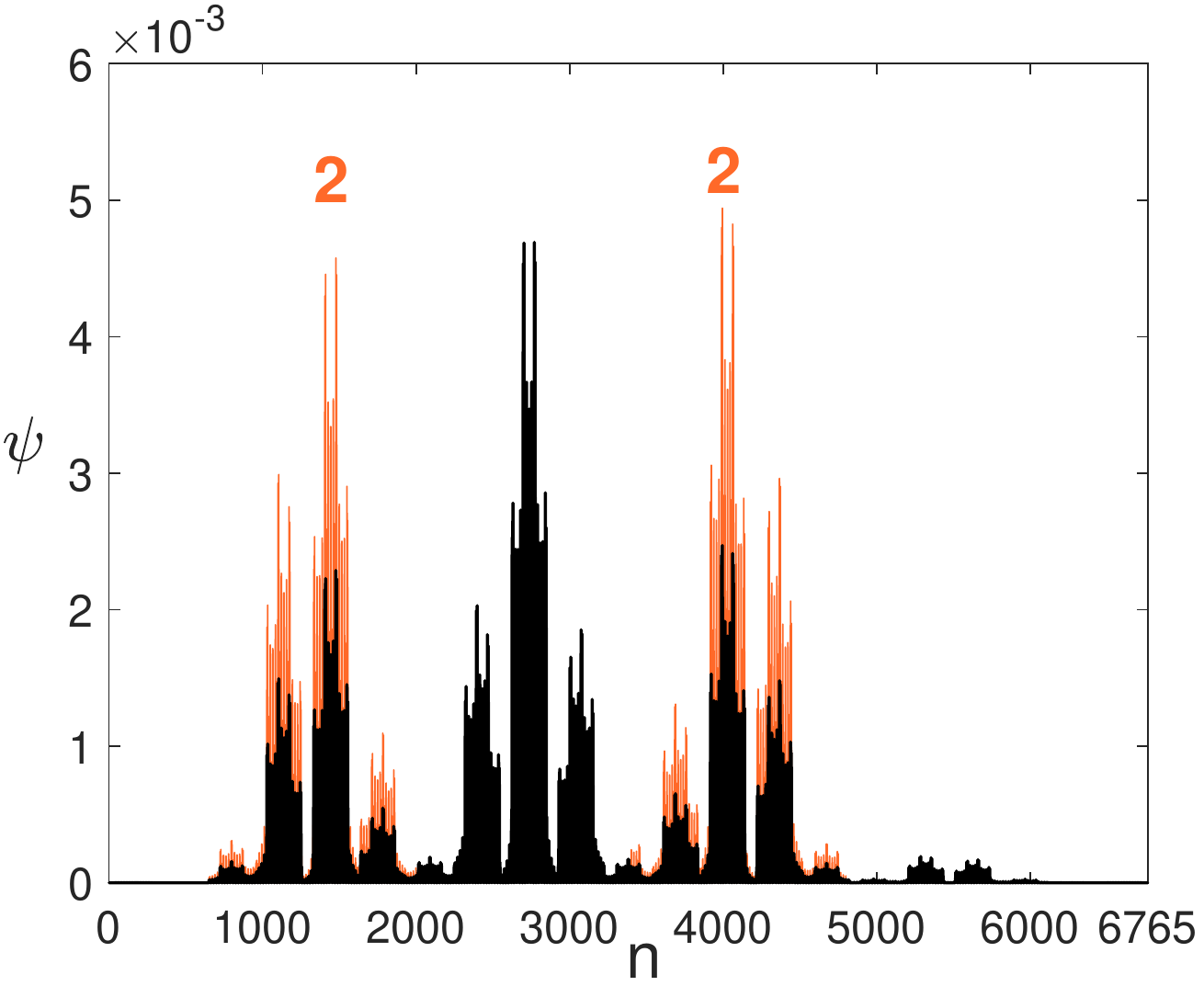}\\
  \caption{(Color online) The black curve represents wave function of $E=0.5$ with the parameter $V=1$. The red curves represent three wave function peaks after magnifying. It clearly shows that the scaled two smaller peaks are very similar to the largest peak after twice magnification. The total number of sites is set to be $L=6765$.}
\label{fig2}
\end{figure}

\begin{figure}
  \centering
  \includegraphics[width=0.5\textwidth]{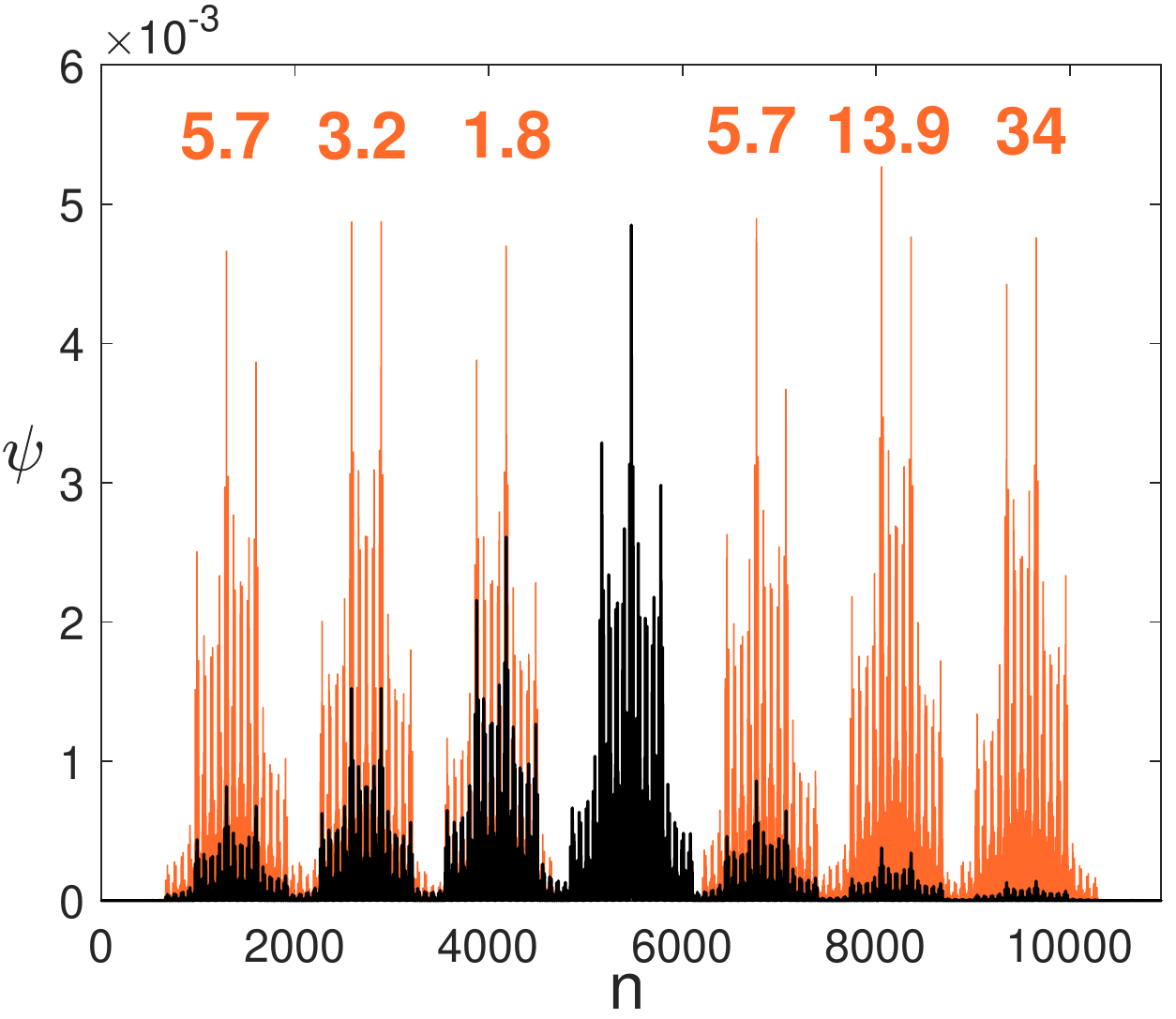}\\
  \caption{(Color online) The black curve represents wave function of $E=-1$ with the parameter $V=1$. The red curves represent three wave function peaks after magnifying. The scaled multiples of three left small peaks are 5.7, 3.2 and 1.8 in turn; the scaled multiples of three right small peaks are 5.7, 13.9 and 34 in turn. The total number of sites is set to be $L=10946$.}
\label{fig3}
\end{figure}

\end{document}